# Low Mass Standard Model Higgs Limit at the Tevatron

J. Keung
*University of Toronto, Toronto, ON, Canada*

The searches for the Standard Model (SM) Higgs Boson at the Fermilab Tevatron by the CDF and DØ experiments are presented. Their state of the art techniques, including maximizing Higgs signal acceptance, reducing background through b-jet ID, and with Multi-Variate discrimination between signal and background, are elucidated. The two experiments are able to achieve a sensitivity of three to five times SM cross section ($\sigma_{SM}$) at the benchmark mass point of $m_H$=115 GeV/c$^2$ using the main search channels WH→lvbb, ZH→vvbb, and ZH→llbb, and on combining all the channels from CDF and DØ, the observed (expected) limit is 1.56 (1.45) x $\sigma_{SM}$. The present expected limit is 1.8 x $\sigma_{SM}$ or below for the entire low mass range, and sensitivity projections at present anticipate in Tevatron Run II a 3$\sigma$ sensitivity achievement for $m_H$=115 GeV/c$^2$.

## 1. INTRODUCTION

The Standard Model (SM) is a very good theory that has made many predictions which have been successfully verified to incredible precision experimentally. But it is still an incomplete theory. One shortcoming of this theory is that it cannot explain the different masses of particles. One proposal to ameliorate this suggests that mass is not an intrinsic property of particles, but that mass arises from the interaction with a yet unseen field, the Higgs field, via the particle called the Higgs boson.

At present, the existence of a Higgs boson is the largest unresolved problem in the Standard Model. Standard electroweak theory uses a single fundamental scalar particle, the Higgs boson, to motivate the spontaneous electroweak symmetry breaking [1], which is needed to explain how the masses of the W and Z bosons arise.

The Tevatron Run II program at Fermilab, located in Batavia, Illinois, USA, has had the capability of producing it since 2001. But unfortunately, its production is predicted to be so rare, that only one Higgs boson is produced for approximately every $10^{11}$ interactions carried out. This poses a difficult challenge to the scientists who want to identify a set of Higgs bosons for studies with the CDF[2] and DØ[3] detectors.

## 2. SEARCH FOR THE HIGGS BOSON

### 2.1. Indirect Searches with Precision Measurements

The Higgs boson enters in Standard Model interactions via radiative corrections, and by measuring very accurately the masses of the top quark ($M_t$) and the W boson ($M_W$), we can constrain the Higgs boson mass. Using data from all precision electroweak measurements together to constraint the Higgs boson mass, we have a 95% C.L. upper limit of $M_H$ < 154 GeV/c$^2$.[4]

### 2.2. Direct Searches at LEP

The Large Electron Positron Collider (LEP) at CERN operated at center of mass energies up to 209 GeV until November 2000. The experimentalists searched for evidence of a Higgs boson directly produced from electrons and positrons collisions. Their search resulted in a constraint of $m_H$ > 114.4 GeV/c$^2$ at 95% C.L. [5].

### 2.3. Direct Searches at Tevatron

The Fermilab Tevatron has been operating at a center of mass energy of 1.96 TeV since 2001. It is expected to deliver 12 fb$^{-1}$ by the end of Run II in 2011. Figure 1 shows the Tevatron production cross section for several processes as a function of input Higgs boson mass. The processes with the largest production cross section are gluon fusion (gg→H) and associated production with W/Z boson (qq→WH/ZH).

Figure 2 shows the branching fractions for several decay modes as a function of Higgs boson mass. The dominant decay mode changes at $M_H$ = 135 GeV/c$^2$ from a pair of bottom quarks (H→bb) to a pair of W bosons (H→WW). The methods for the direct searches for the Higgs boson at the Tevatron are separated into two categories according to the dominant Higgs boson decay modes: low-mass ($M_H$ < 135 GeV/c$^2$) searching for the decay into a bottom quark pair, and high-mass ($M_H$ > 135 GeV/c$^2$) searching for the decay into a W boson pair.

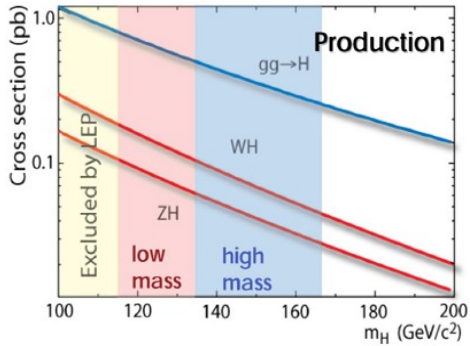
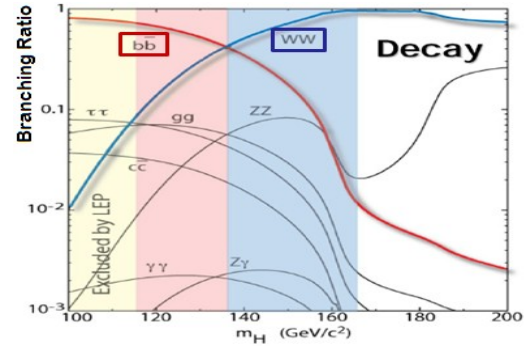

Figure 1: Tevatron SM Higgs boson production cross sections [6].   Figure 2: SM Higgs boson branching ratios [7].

### 2.4. Direct Searches in the Low-Mass Category at the Tevatron

In the low-mass category the dominant mode of the Higgs boson decay is the bottom quark pair. Pairs of bottom quarks are unfortunately produced much more frequently by other processes (10$^6$x) at the Tevatron. However, with the associated production with W/Z boson (qq→WH/ZH), even though it is rarer than the production of a Higgs boson alone, the rate of other processes involved in the associated production of bottom quarks pairs is reduced much more. This gives the associated production channels a higher signal-to-background ratio.

### 3. LOW MASS HIGGS SEARCH STRATEGY

Low mass Higgs search strategies are three fold. Firstly, we maximize signal acceptance to try to collect every last Higgs event produced. Secondly, we reduce background by selecting the b-jet decays. Finally, we use Multi-Variate(MV) tools to further discriminate between signal and background.

### 3.1. Maximizing Signal Acceptance

We loosen event selection to collect more signal events, but in doing so we collect also background events. In order to have significant gain, we must reject the background. For example in the ZH→llbb channel at CDF, a loosened lepton selection is used to gain signal acceptance, and with it a lot of background. However, by using a MuonID Neural Network, most of the background is removed while preserving most of the signal (see Fig. 3).

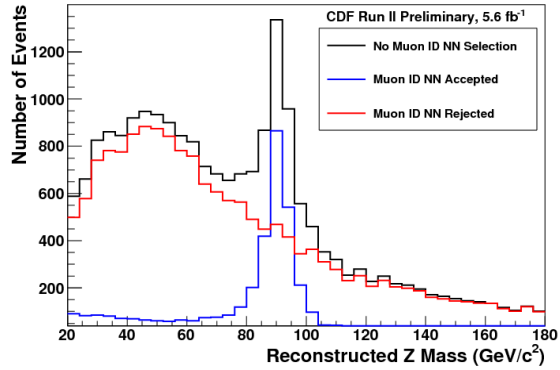

Figure 3: Black line shows the additional acceptance obtained from loosening the muon identification, blue line shows the improvement in purity from using a MuonID Neural Network.

## 3.2. Reduce background through b-jet ID

Since the Higgs decays mostly into a pair of b-jets while most of the background are not, identifying the b-jets improves the signal to background ratio (Figure 4).

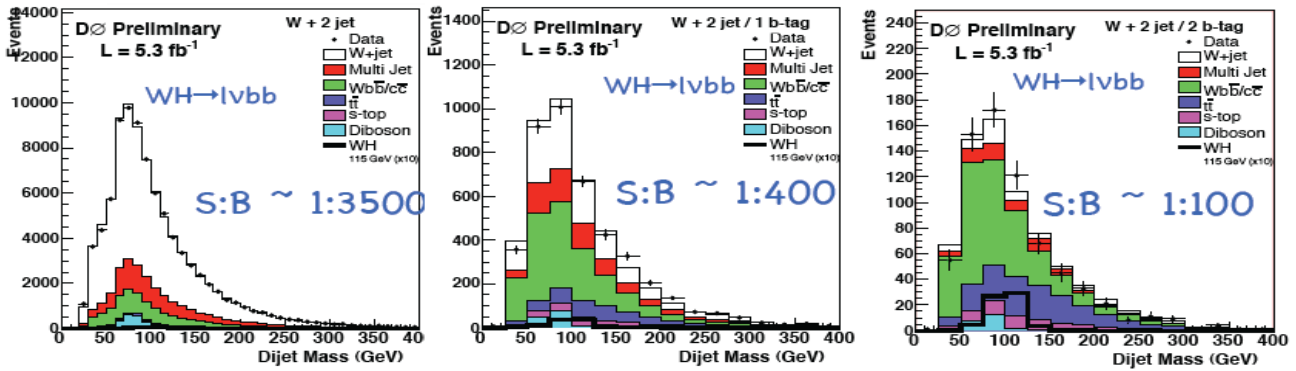

Figure 4: Zero b-tags (left), one b-tag (middle), 2 b-tags (right). S:B ratio improves with additional b-tags.

## 3.3. Multi-Variate (MV) Discrimination between Signal and Background

MV combines multiple discriminants into a stronger one, and is able to improve analyses by ~20% with respect to leading two variables. Common MV Discriminants include Artificial Neural Network (NN), Boosted Decision Trees (BDT), Random Forests (RF), and Matrix Element Probabilities (ME). They have similar performance.

It is important to note that the ~pb processes such as Single top[8],[9], WW/WZ (in lvqq final state)[10],[11] have been observed with MV tools. One caveat however, is that our recent primary sensitivity gains don't come from MV, they mainly come from improved signal acceptance

## 4. MAIN LOW MASS HIGGS SEARCH CHANNELS

The searches for the Higgs boson in the low-mass category can be divided into three channels depending on the mode of decay of the associated W or Z boson. A fourth channel using the direct Higgs production contributes to the sensitivity in the low-mass category, but is instead considered a high-mass channel because it is the main contributor to the sensitivity in the high-mass category. The yields of these channels are shown in Figure 5 and Table 1.

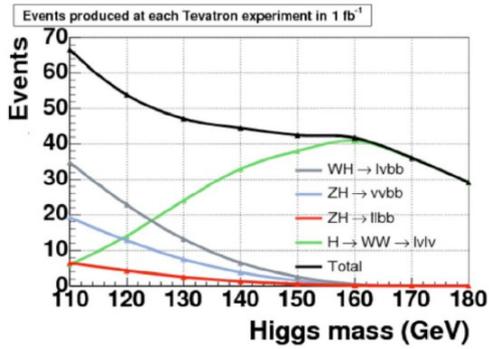

| Production +Decay Channel | events/fb @ 115 GeV/c$^2$ | events/fb @ 165 GeV/c$^2$ |
|---|---|---|
| WH→lvbb | 28 | 0.1 |
| ZH→vvbb | 16 | 0.07 |
| ZH→llbb | 5 | 0.02 |
| H→WW→lvlv | 9 | 38 |

Table 1: Main Higgs Production Channel Yields, excluding trigger/acceptance/selection efficiencies.

Figure 5: Number of expected Higgs events produced as a function of Higgs mass.

### 4.1. WH→lvbb

If the Higgs boson is produced in association with a W boson, then the final state searched for is WH→lvbb, where the W decays into an electron or muon and a neutrino. The decay of W into a tau which then decays into an electron or muon is also included, but the hadronic decay modes are not. Including only the electron or muon and the leptonic decays of tau, 28 events ($m_H$ = 115 GeV/c$^2$) are expected to be produced in this channel for each fb$^{-1}$ of integrated luminosity [7],[12]. The inclusion of using the hadronic decay modes of tau is an area of active development.

### 4.2. ZH→llbb

If the Higgs boson is produced in association with a Z boson, then the final state searched for is ZH→llbb, where the Z decays into a pair of electrons or muons or taus (including the leptonic decays of the tau only). This provides the cleanest signature, since the background processes rarely produce lepton pairs. The decays of the Z boson into a pair of taus are not considered if either tau then decays hadronically, since their identification is more difficult. Approximately 5 events ($m_H$ = 115 GeV/c$^2$) are expected to be produced in this channel for each fb$^{-1}$ of integrated luminosity [7],[12].

### 4.3. VH→Met bb

Be it either the W or Z boson with which the Higgs boson is produced in association, sometimes the leptonic decay products of the W or Z eludes direct identification. Sometimes they fail identification criteria or escapes detection by passing through inactive regions of the detector, or for the Z boson decaying via Z→vv and escaping detection in the face of active regions as well. In these cases we can detect their presence indirectly, from a large momentum imbalance in the transverse plane. Approximately 30 events ($m_H$ = 115 GeV/c$^2$) are expected to be produced in this channel for each fb$^{-1}$ of integrated luminosity [7],[12].

### 4.4. H→WW→lvlv

The second highest decay fraction at the low mass category is H→WW, and 9 events ($m_H$ = 115 GeV/c$^2$) are expected to be produced in this channel for each fb$^{-1}$ of integrated luminosity [7],[12]. But H→WW is the dominant decay mode for the high mass category, where 38 events ($m_H$ = 165 GeV/c$^2$) are expected to be produced in this channel for each fb$^{-1}$ of integrated luminosity [7],[12]. This channel is discussed in more detail in the Tevatron high mass Higgs Limit contribution.

### 4.5. Results from main low mass Higgs search channels

Table 2 shows the observed and expected 95% CL limits for the main Tevatron low mass Higgs search channels.

| 95% CL Limit for $m_H$=115 GeV/c$^2$ | Observed (Expected) [x $\sigma_{SM}$] |
|---|---|
| CDF lvbb Matrix Element 5.6 fb$^{-1}$ [13] | 3.6 (3.5) |
| CDF lvbb Neural Network 5.7 fb$^{-1}$ [14] | 4.5 (3.5) |
| DØ lvbb Random Forest 5.3 fb$^{-1}$ [15] | 4.1 (4.8) |
| CDF Metbb Neural Network 5.7 fb$^{-1}$ [16] | 2.3 (4.0) |
| DØ Metbb Decision Tree 5.5 fb$^{-1}$ [17] | 3.4 (4.2) |
| CDF llbb Neural Network 5.7 fb$^{-1}$ [18] | 6.0 (5.5) |
| DØ llbb Random Forest 5.2 fb$^{-1}$ [19] | 8.0 (5.7) |

Table 2: Observed and expected 95% CL limits for the main Tevatron low mass Higgs search channels.

## 5. LOW MASS HIGGS CHANNELS COMBINATION

With the current state of the art analysis techniques and present dataset, no single channel has significant sensitivity to exclude the Low Mass Higgs. Therefore we combine the many channels together to maximize sensitivity.

### 5.1. CDF Combination

Figure 6 shows the CDF combined SM Higgs 95% CL limits for different possible Higgs masses. The observed (expected) limit is 1.79 (1.90) times Standard Model cross section for $m_H$=115 GeV/c$^2$.

### 5.2. DØ Combination

Figure 7 shows the DØ combined SM Higgs 95% CL limits for different possible Higgs masses. The observed (expected) limit is 2.65 (2.31) times Standard Model cross section for $m_H$=115 GeV/c$^2$.

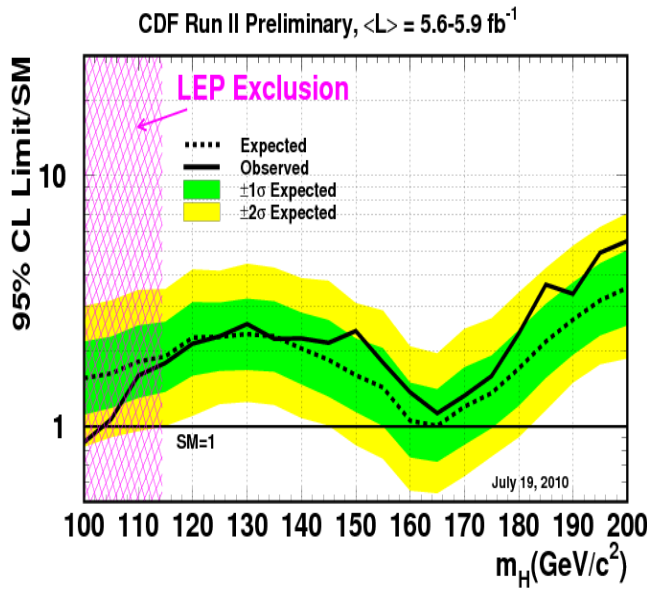
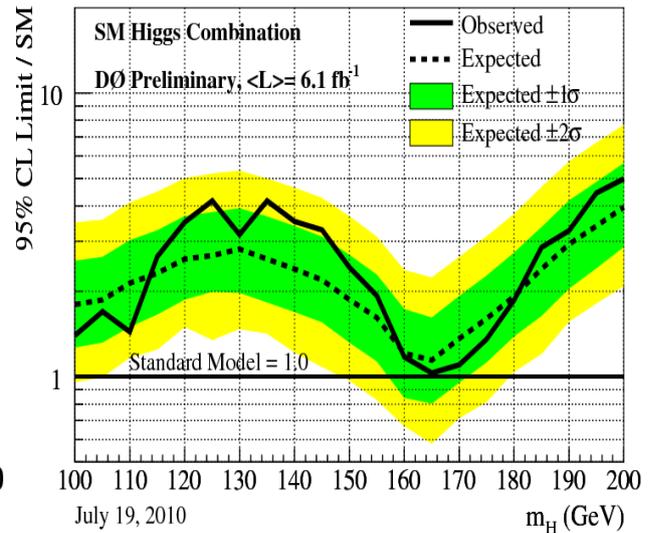

Figure 6: CDF combined SM Higgs 95% CL limits.          Figure 7: DØ combined SM Higgs 95% CL limits.

## 5.3. Tevatron Combination

Neither CDF nor DØ alone has enough sensitivity to exclude the Low Mass Higgs, so we combine the two experiments together to maximize sensitivity. Figure 8 shows the Tevatron combined SM Higgs 95% CL limits for different possible Higgs masses. For $m_H$=115 GeV/c$^2$, the observed (expected) limit is 1.56 (1.45) times Standard Model cross section. We note that the expected limit is 1.8 x $\sigma_{SM}$ or below for entire low mass range, and that we are beginning to exclude some masses below the direct LEP limits.

## 6. OUTLOOK

Presently CDF and DØ each have about 8 fb$^{-1}$ recorded, and is each projected to collect ~2 fb$^{-1}$ more per year. The Tevatron is scheduled to run at least one more year, thus one can expect each experiment to collect about 10 fb$^{-1}$. Based on the sensitivity projection in Figure 9, we can expect to achieve 3σ sensitivity at $m_H$=115 GeV/c$^2$.

The improvements projected to be achieved include the migration of existing improvements across channels, the expansion of e/µ selection, the inclusion of final states with taus, improvements in b-jet identification, and improvements in jet energy resolution.

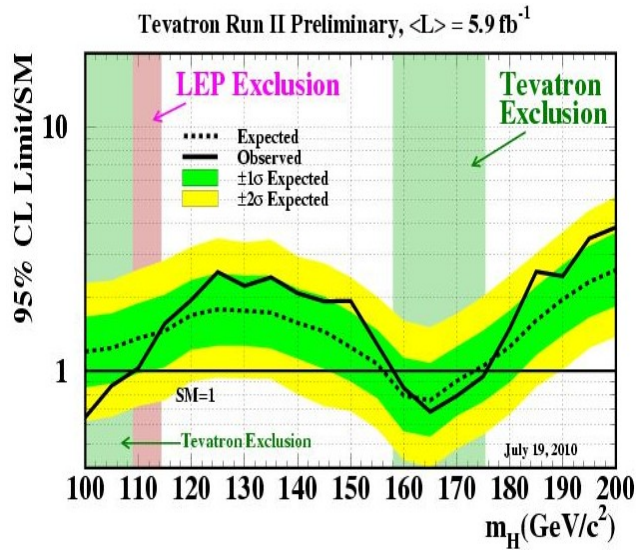 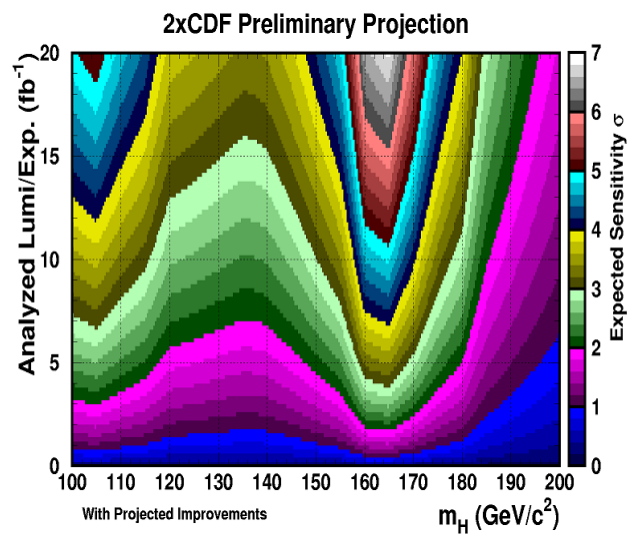

Figure 8: Tevatron combined SM Higgs 95% CL limits.

Figure 9: Tevatron sensitivity projection made by CDF, assuming same sensitivity from DØ.

## 7. CONCLUSIONS

The Tevatron combined 95% CL limit for SM Higgs with $m_H$=115 GeV/c$^2$ is observed (expected) 1.56 (1.45) x $\sigma_{SM}$. The Tevatron is already excluding some masses below the direct LEP limits, and with projected improvements, is expected to be able to exclude SM Higgs (if it does not exist) over the entire low mass region. The coming years should prove to be exciting in the search for the Low Mass Higgs.